\documentclass[epjCONF,columns]{svjour} 
\usepackage{graphicx}
\usepackage{amsmath}
\usepackage[varg]{txfonts} 
\usepackage[latin1]{inputenc}
\usepackage{color}
\session-title{Hadron Collider Physics Symposium 2011}

\def\mrm{\mathrm}
\def\ra{\rightarrow}

\newcommand{\afb}{\ensuremath{A_{FB}}}
\newcommand{\fl}{\ensuremath{F_{L}}}

\newcommand{\GeV}{\ensuremath{\mathrm{Ge\kern -0.1em V}}}
\newcommand{\MeV}{\ensuremath{\mathrm{Me\kern -0.1em V}}}

\newcommand{\fb}{\ensuremath{\mathrm{fb}^{-1}}}

\newcommand{\bs}{\ensuremath{B^{0}_{s}}}
\newcommand{\bd}{\ensuremath{B^0_d}}
\newcommand{\bsd}{\ensuremath{B^{0}_{s,d}}}
\newcommand{\bu}{\ensuremath{B^+}}
\newcommand{\jp}{\ensuremath{J/\psi}}
\newcommand{\mm}{\ensuremath{\mu^{+}\mu^{-}}}

\newcommand{\bmmks}{\ensuremath{\bd\ra\mm K^{*0}(892)}}
\newcommand{\bpmmk}{\ensuremath{B^+\ra\mm K^+}}
\newcommand{\bmmphi}{\ensuremath{\bs\ra\mm \phi}}
\newcommand{\bpmmksp}{\ensuremath{B^+\ra\mm K^{*+}}}
\newcommand{\bmmksComb}{\ensuremath{B\ra\mm K^{*}}}
\newcommand{\lambdabmmlambda}{\ensuremath{\Lambda_b\ra\mm \Lambda}}
\newcommand{\bmmkshort}{\ensuremath{\bd\ra\mm K_s}}

\newcommand{\bsmm}{\ensuremath{\bs\ra\mm}}
\newcommand{\bdmm}{\ensuremath{\bd\ra\mm}}
\newcommand{\bsdmm}{\ensuremath{\bsd\ra\mm}}
\newcommand{\jpmm}{\ensuremath{\jp\ra\mm}}
\newcommand{\bjk}{\ensuremath{\bu\ra\jp K^{+}}}

\newcommand{\brbsmm}{\ensuremath{\mathcal{B}(\bsmm)}}
\newcommand{\brbdmm}{\ensuremath{\mathcal{B}(\bdmm)}}

\newcommand{\brbsmmo}{\ensuremath{4.0\times 10^{-8}}}

\newcommand{\Mmm}{\ensuremath{M_{\mm}}}

\newcommand{\cdf}{CDF~II}

\newcommand{\pt}{\ensuremath{p_{T}}}
\newcommand{\hb}{\ensuremath{H_{b}}}
\newcommand{\AT}{\ensuremath{\mathrm{A}^{(2)}_{\mathrm{T}}}}
\newcommand{\AIm}{\ensuremath{\mathrm{A}^{\mathrm{Im}}}}

\begin{document}
\title{Rare $B$ Meson Decays at the Tevatron}
\author{Walter Hopkins\inst{1}\fnmsep\thanks{\email{whh0407@FNAL.GOV}} }
\institute{Cornell University}
\abstract{ 
  Rare $B$ meson decays are an excellent probe for beyond the Standard Model physics. 
  Two very sensitive processes are the $b\to s\mu^{+}\mu^{-}$ and \bsdmm\ decays. We report recent results at a center of mass energy of $\sqrt{s} = 1.96$~TeV from
 \cdf\  using 7 \fb\ at the Fermilab Tevatron Collider. 
} 
\maketitle

%


\section{Introduction: $b\to s\mu^{+}\mu^{-}$ decays}
$b\to s\mu^{+}\mu^{-}$ decays are flavor changing neutral current (FCNC) processes that can only occur through higher order box or penguin amplitudes
in the Standard Model. New physics can be probed by measuring various combinations of their decay rates. One of the most sensitive 
observables is the forward-backward asymmetry of the muons (\afb ) as a function of the squared di-muon mass. 

Decays of interest are \bpmmk, \bmmks, \bmmphi, \bpmmksp, \bmmkshort, \lambdabmmlambda . 
The CDF analysis that uses 6.8 \fb\ of data is described here~\cite{btosmmAsym},~\cite{btosmmBR}.

\subsection{Analysis Method}

\subsubsection{Branching Ratio Measurement}
Branching ratios for the $b\to s\mu^{+}\mu^{-}$ decays are 
measured relative to normalization modes, where the two muons originate from a $J/\psi$ decay. For the event
reconstruction CDF requires two muons with a transverse momentum ($p_T$) greater than either 1.5 GeV/c, 2.0 GeV/c, or 3.0 GeV/c depending on
the muon trigger. The six modes are then reconstructed where the $K^{*0}(892)$ is reconstructed from $K^{*0}(892)\to K^+\pi^-$, the $\phi$ 
from $\phi\to K^+ K^-$, the $K^{*+}$ from $K^{*+}\to K_s\pi^+$, the $K_s$ from $K_s\to\pi^+\pi^-$, and the $\Lambda$ from $\Lambda\to p\pi^-$. 
To avoid contamination from resonant modes
such as the $J/\psi$ and $\psi'$, candidates with di-muon masses near these resonances are rejected.

The events then have to meet loose preselection requirements before an artificial neural network (NN),
which combines multiple discriminating variables into one variable, is applied. Signal is modeled with $p_T$-reweighted 
Pythia signal Monte Carlo simulations (MC).
The reweighing is done by comparing the MC $p_T$ distribution with that of the normalization modes. The background is modeled by sampling 
the $b$ hadron (\hb) mass sideband regions. 

The final signal yield is obtained by an unbinned maximum likelihood fit to the \hb\ mass distribution. The probability distribution function (PDF) of the signal
is parametrized with two Gaussian with different means while the the background PDF is described by a first or second order polynomial. 
Peaking background contributions are subtracted from the fit results for the signal yields. The only significant peaking contribution is cross-talk among 
\bmmks\ and \bmmphi\ which has a $\sim$1\% contribution to the total observed signal MC yields. 
The final branching ratios are calculated as follows:

\begin{equation}\label{eq:brB}
\begin{aligned}
  \mathcal{B}(\hb\to h\mu^{+}\mu^{-}) = \frac{N_{h\mu^{+}\mu^{-}}^{\mrm{NN}}}{N_{J/\psi h}^{\mrm{loose}}}
  \cdot
  \frac{\epsilon^{\mrm{loose}}_{J/\psi h}}{\epsilon^{\mrm{loose}}_{h\mu^{+}\mu^{-}}}
  \frac{1}{\epsilon^{\mrm{NN}}_{h\mu^{+}\mu^{-}}}\cdot
 & \mathcal{B}(J/\psi h) \\
\cdot \mathcal{B}(J/\psi \to\mu^{+}\mu^{-} ),
\end{aligned}
\end{equation}
where \hb\ signifies the $B^{0}_{d}$, $B^+$, $B^{0}_{s}$, or $\Lambda_b$ and $h$ represents $K^{*0}(892)$, $K^+$, $K^{*+}$, $K_s$, $\Lambda$, or $\phi$, 
$N_{h\mu^{+}\mu^{-}}^{\mrm{NN}}$ and $N_{J/\psi h}^{\mrm{loose}}$ are the yields after the optimal NN selection, 
$\frac{\epsilon^{\mrm{loose}}_{J/\psi h}}{\epsilon^{\mrm{loose}}_{h\mu^{+}\mu^{-}}}$ is the relative efficiency of the loose selection cuts, and 
$\epsilon^{\mrm{NN}}_{h\mu^{+}\mu^{-}}$ is the NN cut efficiency on the loosely-selected events. The NN is not applied to 
the normalization mode because the signal/purity and size is sufficient with the loose selection cuts.
The NN cut efficiency are obtained from signal MC.

The three leading systematics are the systematics on the efficiency, $\mathcal{B}(\hb\to J/\psi h)$, and background PDF. 
The main sources within the efficiency systematic errors are the MC reweighing and the the trigger turn on. The total 
systematics for the \bpmmk, \bmmks, \bmmphi, \bpmmksp, \bmmkshort, \lambdabmmlambda\ 
are 5\%, 6\%, 32\%, 8\%, 6\%, and 32\%, respectively. 

\subsubsection{Forward-Backward Asymmetry Measurement}

For the \bmmks\ and \bmmksComb\ (\bmmks\ and \bpmmksp\ combined fit for increased sensitivity) 
decays the forward-backward asymmetry (\afb ), the $K^{*}$ longitudinal polarization (\fl), 
the transverse polarization asymmetry (\AT), as well as the triple product asymmetry of the transverse polarizations (\AIm)
are measured. For \bpmmk\ only \afb\ is measured. The \afb\ and \fl\ angular quantities are extracted from $\cos\theta_{\mu}$, 
the cosine of the helicity angle between the $\mu^{+}$ ($\mu^-$) momentum vector and the opposite of $B$ ($\bar{B}$)
meson momentum vector in the di-muon rest frame, and $\cos\theta_K$, the cosine of the angle between kaon momentum and 
the opposite of the $B$ meson momentum vector in the $K^{*0}(892)$ rest frame. \AT\ and \AIm\ are extracted 
from $\phi$, where the angle $\phi$ is the angle between the two decay planes
of the di-muon pair and $K-\pi$ decay pair. B decay amplitudes are calculated using operator product expansion and Wilson coefficients.  
There are many non-SM predictions for \afb\ from models with different Wilson coefficients \cite{wilson}.

The angular measurements are extracted using an unbinned maximum likelihood fit containing $B$ mass shape signal and background PDF's as well
as signal and background angular shape PDF's. The mass shape PDF's are divided into several di-muon mass bins and are described as 25 bin histograms.
The combinatorial background PDF is taken from the $B$ meson higher sideband. 
The angular acceptances are also described as 25 bin histograms and are derived from phase space signal MC. 

As a control \afb\ and \fl\ are fitted to $B^0_{d}\rightarrow J/\psi K^{*0}(892)$ and \afb\ only to $B^+\rightarrow J/\psi K^+$. This cross check
yielded measurements that were consistent with other measurements. 

The dominant systematic uncertainties for all angular measurements are the signal fraction and $B$ mass shape uncertainties.
This uncertainty is assessed by varying the signal fraction and shape parameters by $\pm1\sigma$ (statistical uncertainty) when 
performing the B mass fit.

\subsection{Results}
\begin{figure}[htb]
  \centering
  \includegraphics[width=1.55in]{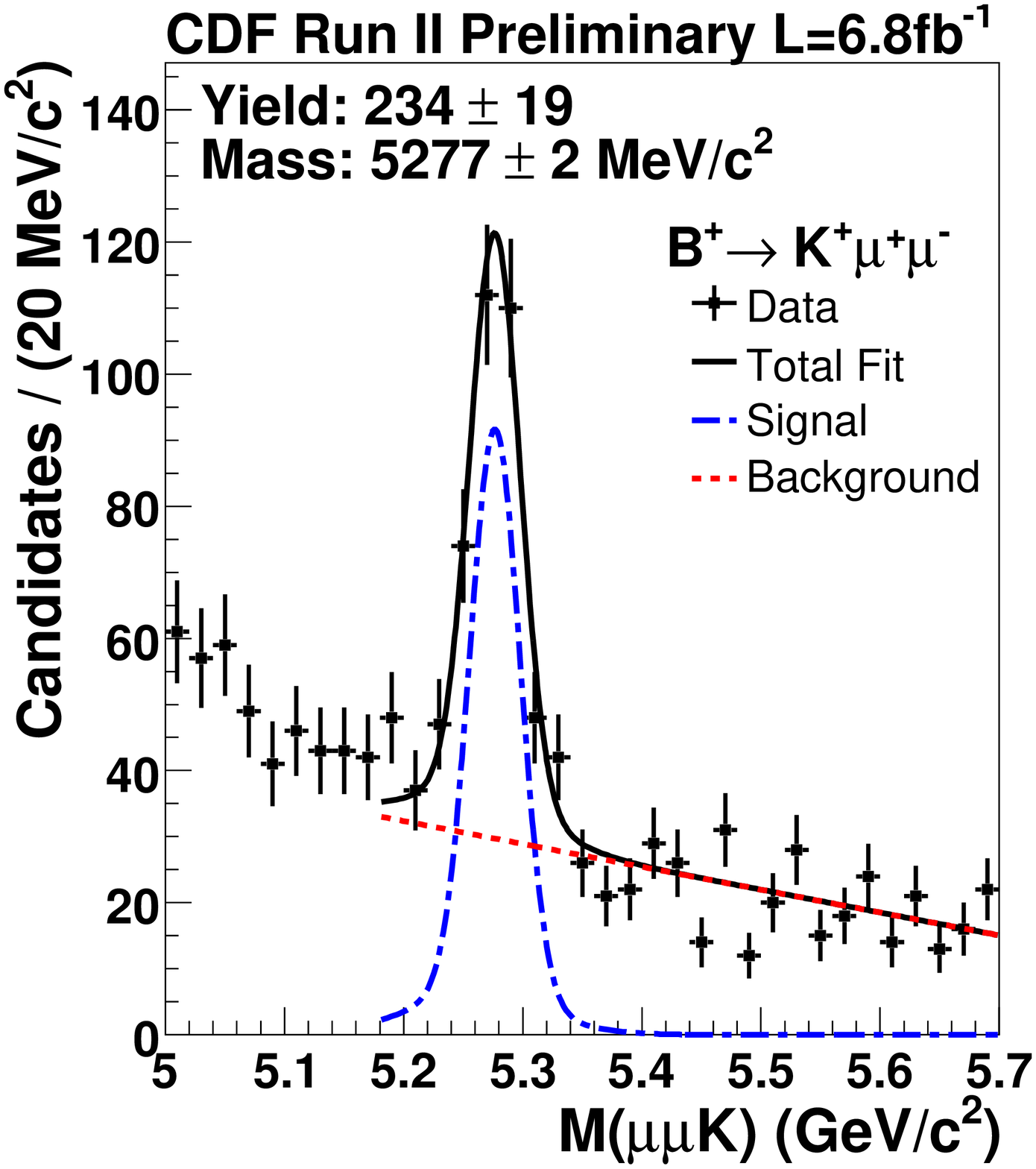}
  \includegraphics[width=1.55in]{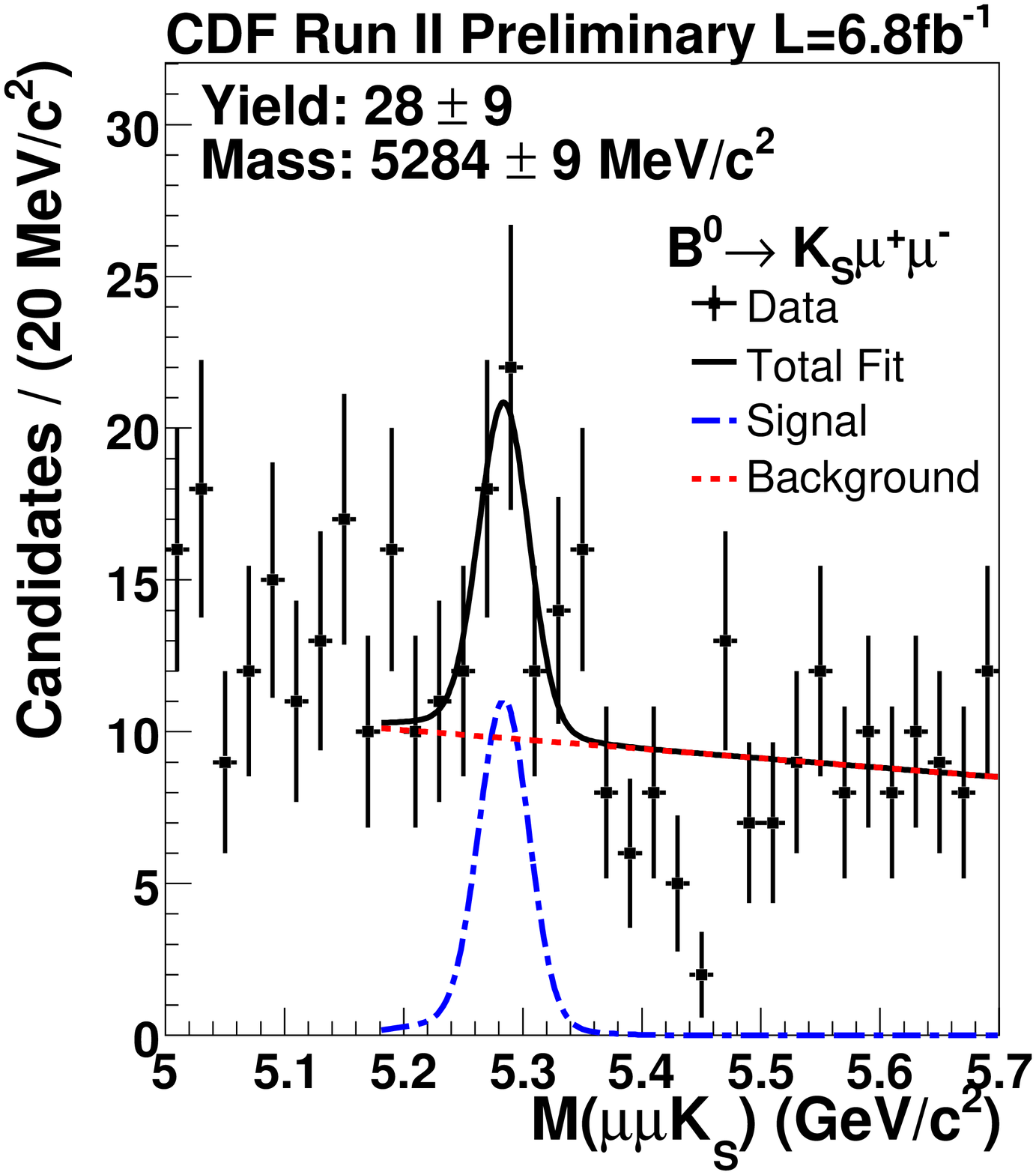}\\
  \includegraphics[width=1.55in]{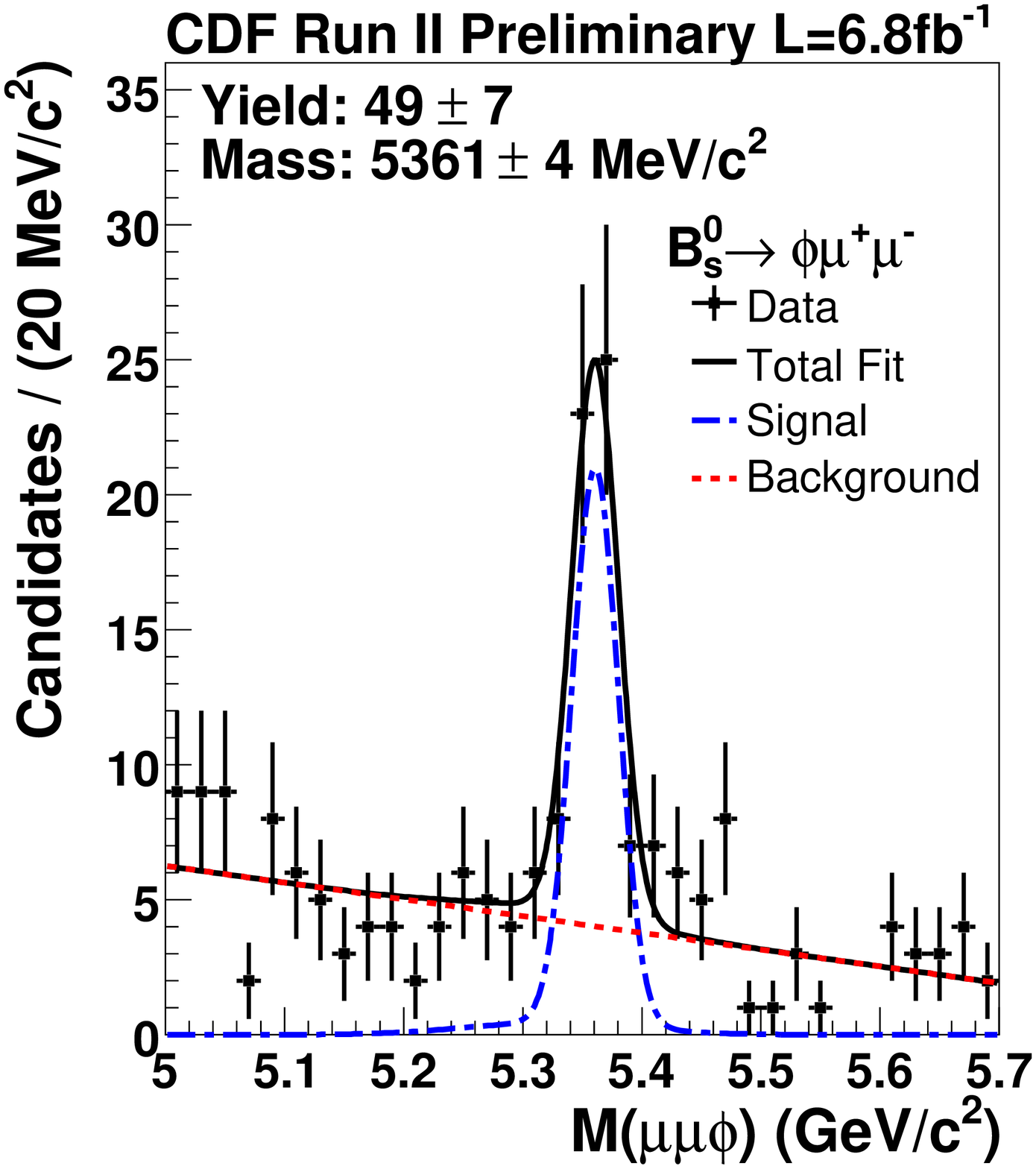}
  \includegraphics[width=1.55in]{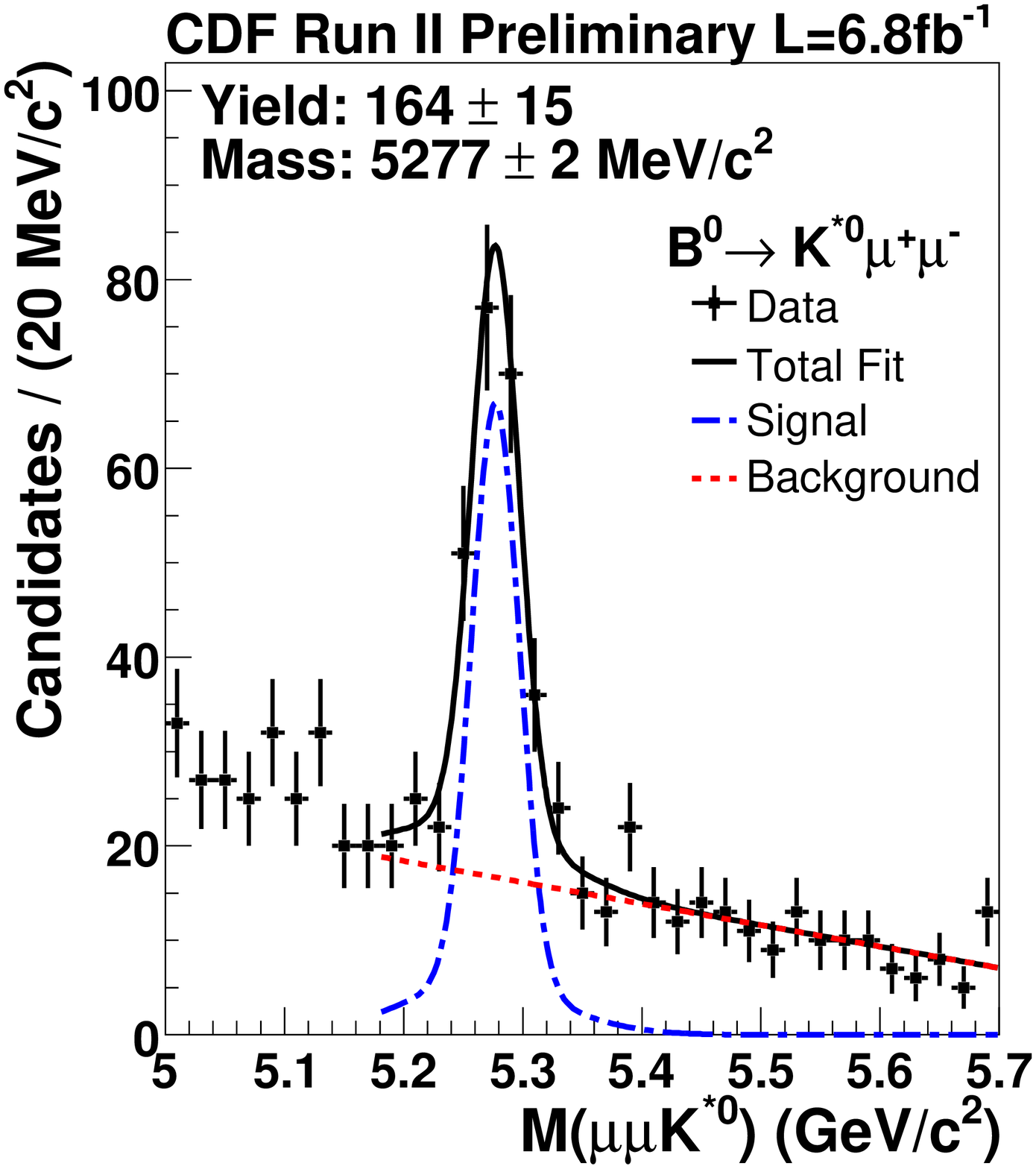}\\
  \includegraphics[width=1.55in]{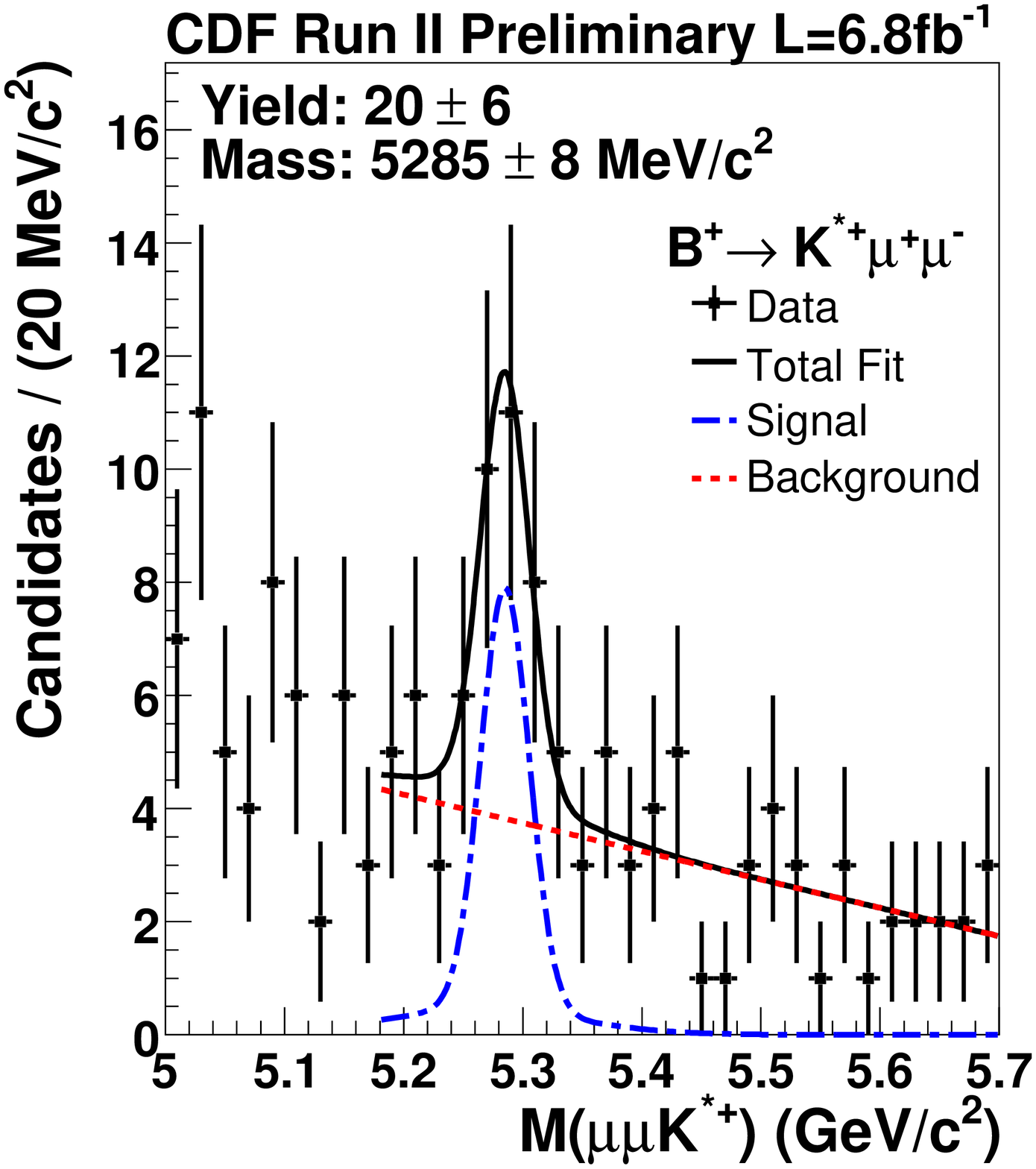}
  \includegraphics[width=1.55in]{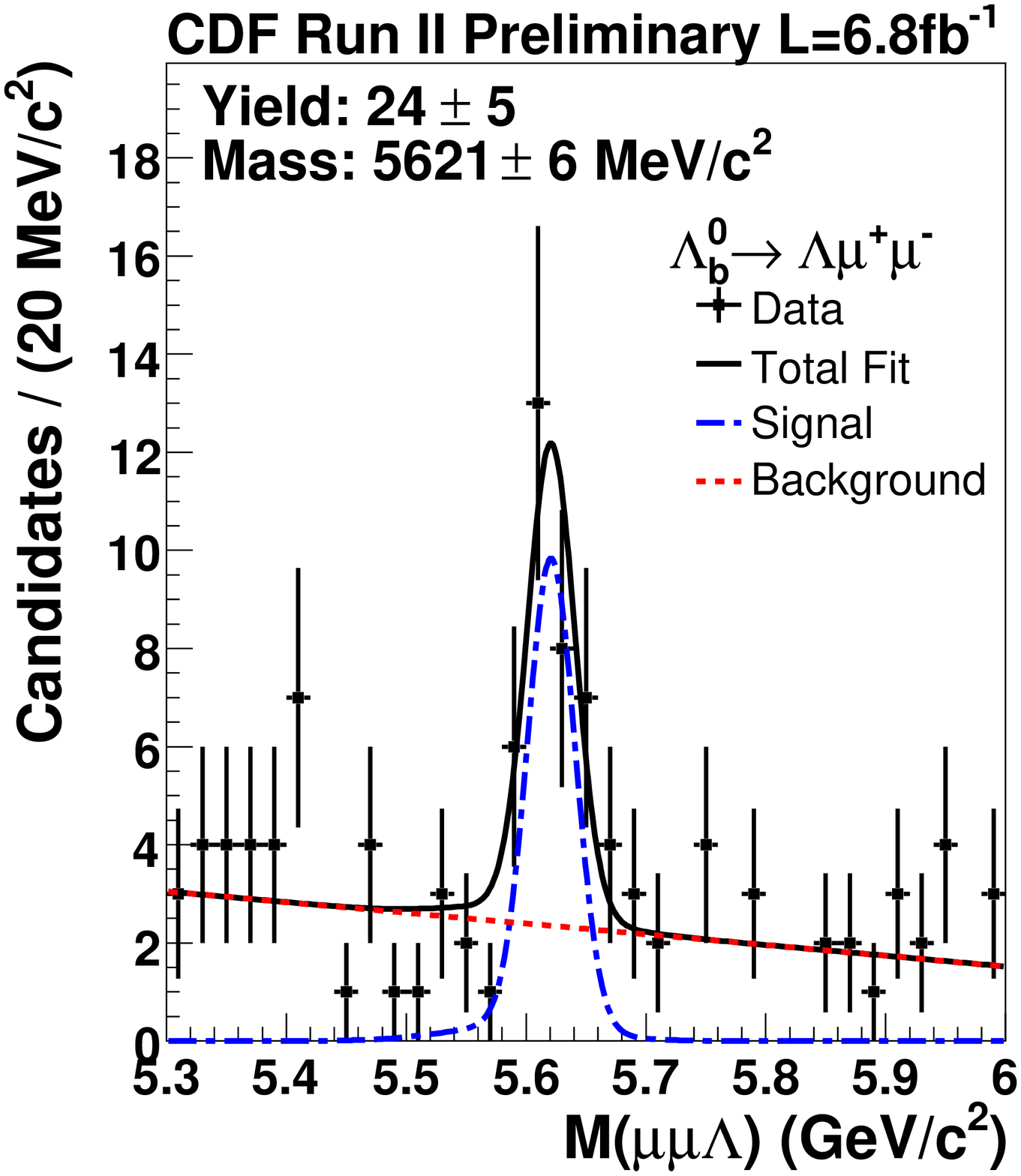}
  \caption{Yields for the various $b\to s \mm$ decays.}
  \label{fig:btosmmYield}
\end{figure}

The resulting yield for the six decays are shown in Figure~\ref{fig:btosmmYield}. All measured branching ratios agree with previous measured 
values as well as theoretical predictions. CDF reports the first observation of \lambdabmmlambda\ with a significance of
$\sim 6\sigma$ and a measured branching ratio of  $\mathcal{B}(\lambdabmmlambda)=(1.73\pm0.42$[stat]$\pm0.55$[syst]$)\times 10^{-6}$.

\begin{figure}[htb]
  \centering
  \includegraphics[width=1.55in]{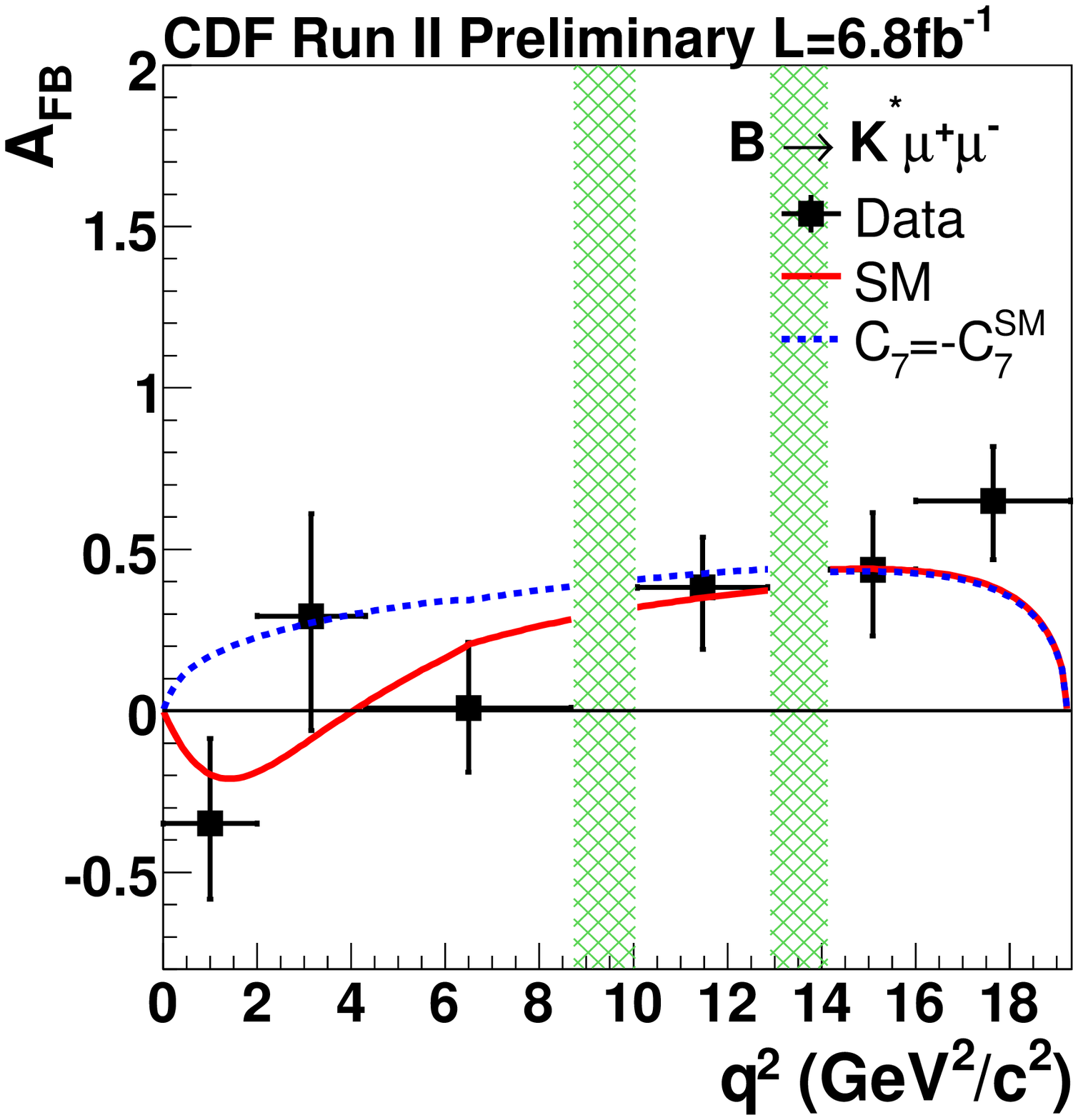}
  \includegraphics[width=1.55in]{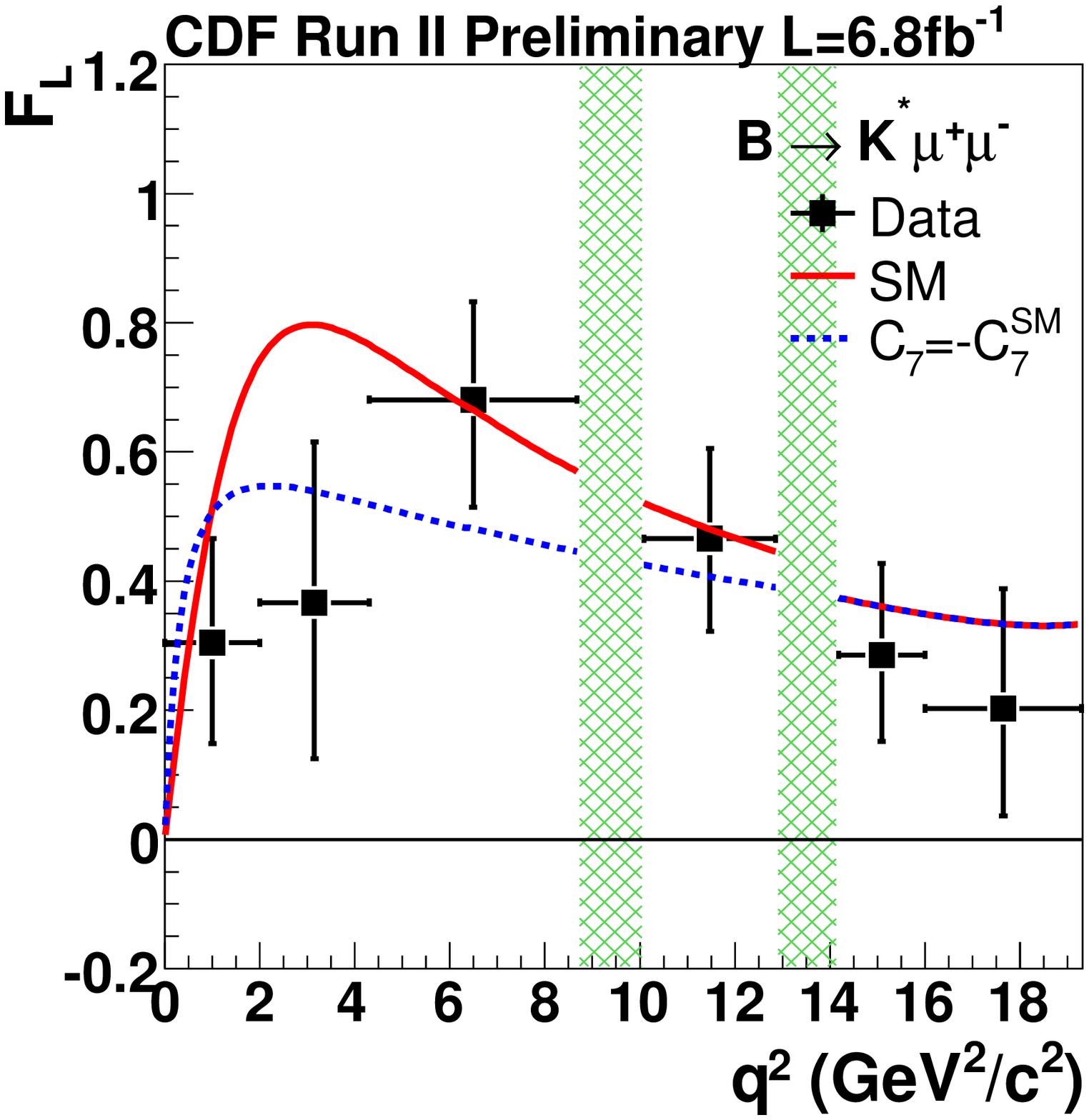}\\
  \includegraphics[width=1.55in]{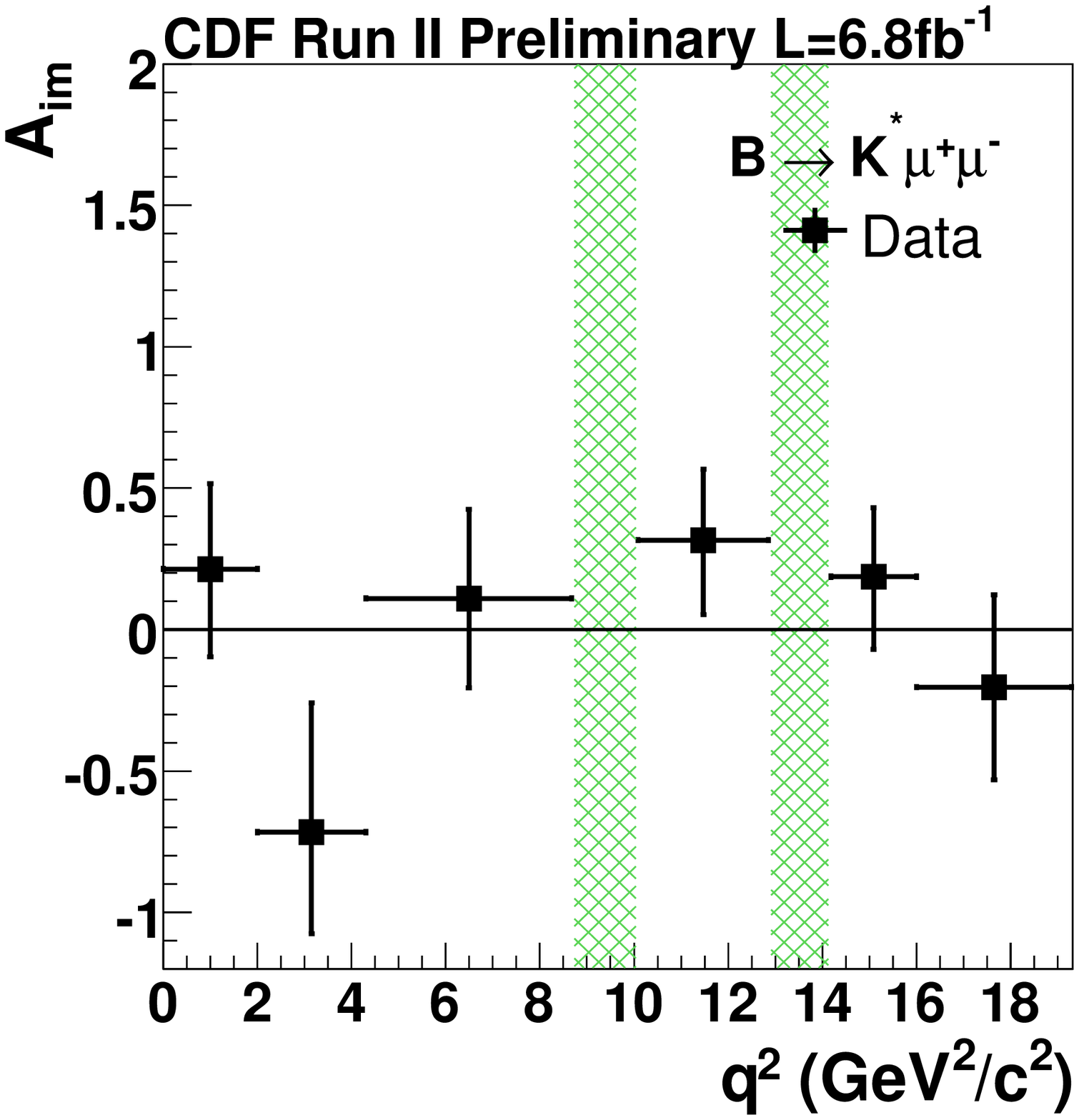}
  \includegraphics[width=1.55in]{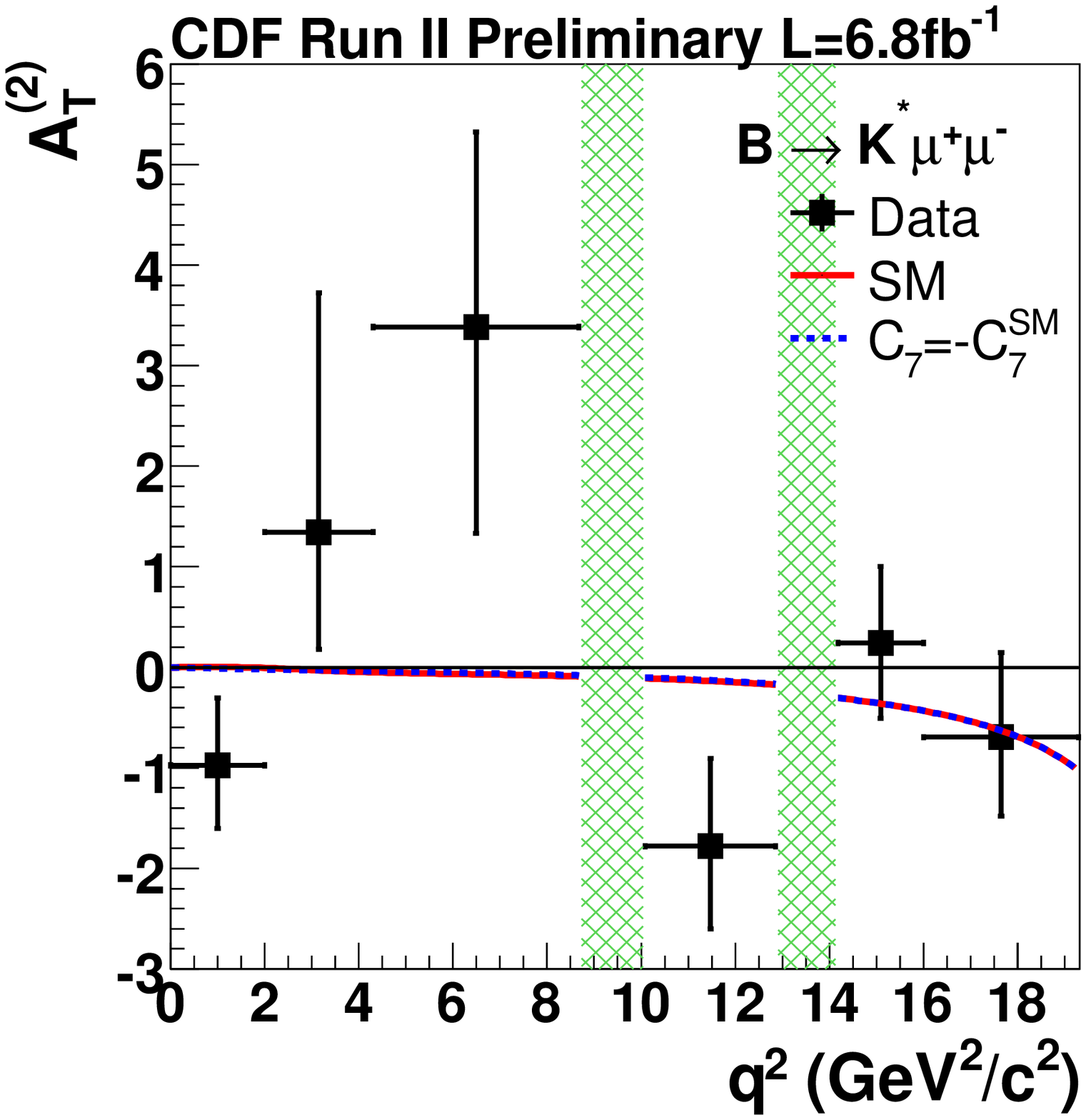}
  \caption{Results of angular measurements for \bmmksComb.}
  \label{fig:ang}
\end{figure}

The results of the angular measurements for the combined \bmmksComb\ fit are shown in Figure~\ref{fig:ang}. 
They are compatible with theory predictions and competitive with the results from the B-factories. 

\section{Introduction: \bsdmm\ decays}

\bsdmm\ are FCNC decays that are highly suppressed by the SM.  The SM predictions for these branching fractions are $\brbsmm = (3.2\pm0.2)\times10^{-9}$ and $\brbdmm = (1.00\pm0.1)\times10^{-10}$~\cite{smbr}.
These predictions are one order of magnitude smaller than the current experimental sensitivity.   Previous bounds from the CDF collaboration, based on 3.7 $\:\fb$ of integrated luminosity, are $\brbsmm < 4.3\times10^{-8}$ and $\brbdmm < 7.6\times10^{-9}$ at $95\%$ C.L.~\cite{cdf9860}.

Enhancements to the rate of \bsmm\ decays occur in a variety of different new-physics models.  In supersymmetric (SUSY) models, new particles can increase \brbsmm\ by several orders of magnitude at large $\tan\beta$.  In the minimal supersymmetric standard model (MSSM), the enhancement is proportional to $\tan^6\!\beta$.  Global analysis including all existing experimental constraints suggest that the large $\tan\!\beta$ region is of interest~\cite{tanb,rparity,dark,chi2}. 

This document describes the current status from CDF. 

\subsection{Analysis Method}

CDF collects opposite sign muon candidate events using di-muon triggers. 

\bjk\ events are collected on the same triggers as a relative 
normalization mode to estimate \brbsmm\ as:
\begin{equation}\label{eq:intro}
\begin{aligned}
  \brbsmm = \frac{N_{\bs}}{\alpha_{\bs}\epsilon^{\mrm{total}}_{\bs}}\cdot
  \frac{\alpha_{B^{+}}\epsilon^{\mrm{total}}_{B^{+}}}{N_{B^{+}}}\cdot 
  &\frac{f_{u}}{f_{s}}\\
\cdot  \mathcal{B}(\bjk)\cdot \mathcal{B}(\jpmm),
\end{aligned}
\end{equation}
where $N_{\bs}$ is the number of candidate \bsmm\ events, $\alpha_{\bs}$
is the geometric and kinematic acceptance of the di-muon trigger for
\bsmm\ decays, $\epsilon^{\mrm{total}}_{\bs}$ is the total efficiency
(including trigger, reconstruction and analysis requirements) for \bsmm\
events in the acceptance, with $N_{B^{+}}$, $\alpha_{B^{+}}$, and
$\epsilon^{\mrm{total}}_{B^{+}}$ similarly defined for \bjk\ decays.
The ratio $f_{u}/f_{s}$ accounts for the different $b$-quark fragmentation 
probabilities and is $(0.402\pm 0.013)/(0.112\pm 0.013) = 3.589 \pm 0.374$, 
including the (anti-)correlation between the uncertainties~\cite{PDG2010}. The final two terms are the relevant branching ratios 
$\mathcal{B}(\bjk)\cdot \mathcal{B}(\jpmm) = 
(1.01\pm0.03)\times10^{-3}\:\cdot\:(5.93\pm0.06)\times10^{-2}
=(6.01\pm0.21)\times10^{-5}$~\cite{PDG2010}.

The CDF analysis described is also sensitive to \bdmm\ decays.  The value of \brbdmm\ 
is estimated from Equation~\ref{eq:intro} substituting \bd\ for \bs, and
changing $f_u / f_s$ to $f_u / f_d = 1$.  All other aspects are the
same as the \bsmm\ search except where noted below.
 
The analysis is done by first estimating the acceptances and efficiencies, then
creating a multivariate discriminant for background rejection. This discriminant is
optimized a priori with Pythia signal MC and data mass sideband events and validated with the \bjk\ normalization mode.
Before the discriminant is applied the events are required to pass baseline cuts that consist of loose requirements on
muon ID, vertexing related variables, muon \pt, and di-muon \pt.
The background is then estimated, which has two sources: combinatorial background and peaking background (B$\rightarrow h^+h^-$).
Finally, when the background is well understood the di-muon mass signal region is unblinded. 

\subsection{Signal and Background Properties}
The signal candidates are fully reconstructed events with a secondary vertex due to the long lifetime ($\sim450\mu$m) of the \bs\ meson. 
Signal events will have a primary-to-secondary vertex vector that is aligned with the \bs\ candidate momentum vector.
Another property of signal events that is unique is that they are very isolated (with few tracks near the muon tracks) due to the hard B fragmentation.

Background events tend to be partially reconstructed and be shorter lived than signal. 
They also have a softer $p_T$ spectrum, higher activity of tracks, and misaligned primary-to-secondary vertex and momentum vectors. 
The combinatorial background consists of sequential semi-leptonic decay ($b\rightarrow c\mu^- X \rightarrow \mu^+\mu^-X$), 
 double semi-leptonic decay ($bb\rightarrow\mu^-\mu^+X$), continuum di-muon events, as well as fake+$\mu$ and fake+fake events. 
The expected number of combinatorial background events in the signal window 
is estimated by extrapolating the number of events in the sideband regions 
to the signal window using a first order polynomial fit. 

The peaking background from two-body hadronic $B$ decays is also evaluated.
These backgrounds are about a factor ten smaller in the \bs\ signal region than the combinatorial background and  need to be estimated separately.
The dominant contributions to this source of background are the decays of $B^{0}_{s}$ and $B^0$ to $h^+ h^{\prime -}$ final
states, where $h$ or $h^\prime$ can be either $\pi^\pm$ or $K^\pm$ and are misreconstructed (fake muons).

\subsection{Analysis Improvements and Signal Discrimination}
\begin{figure}[htb]
  \centering
  \includegraphics[height=1.5in]{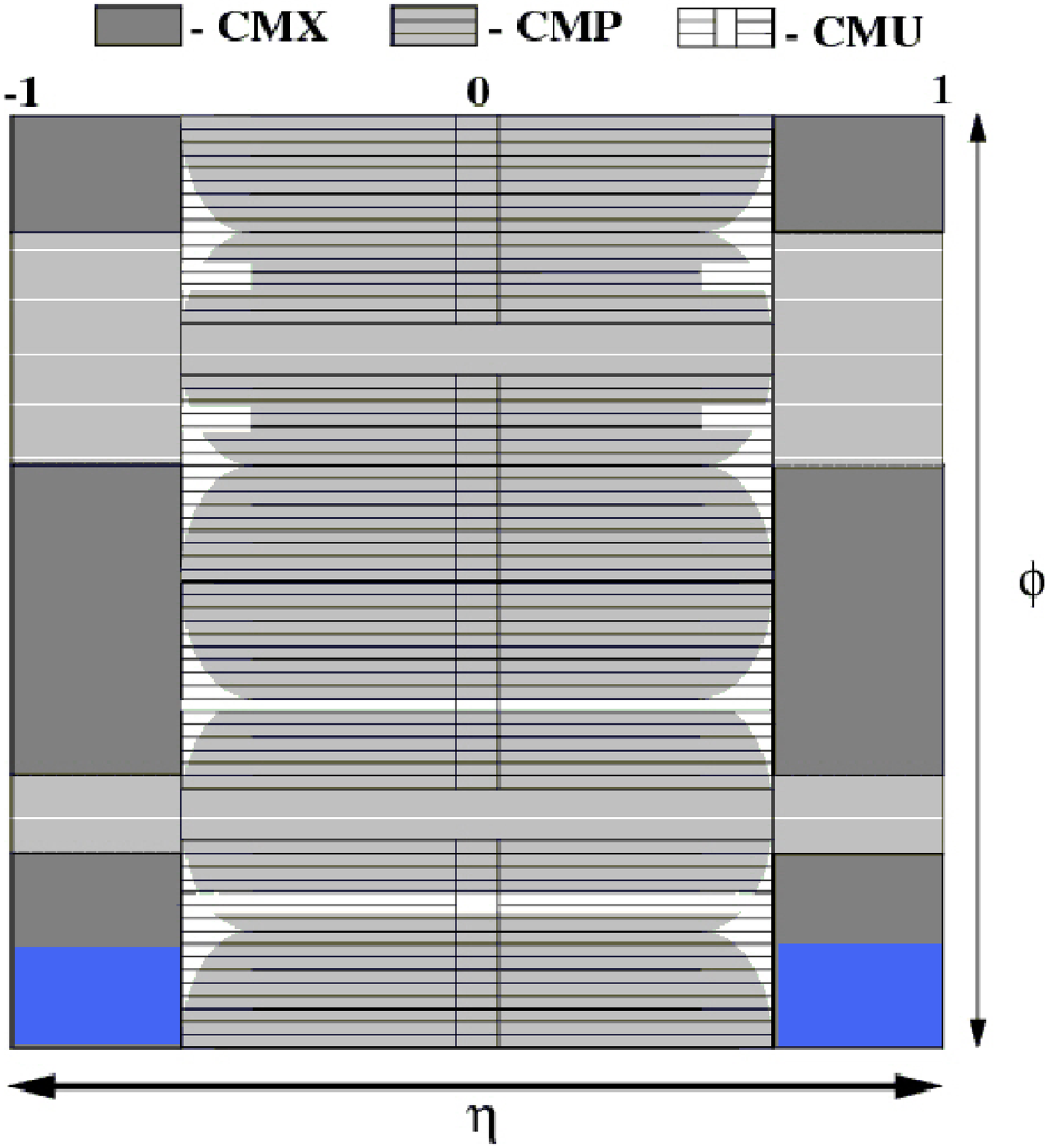}
  \includegraphics[height=1.5in]{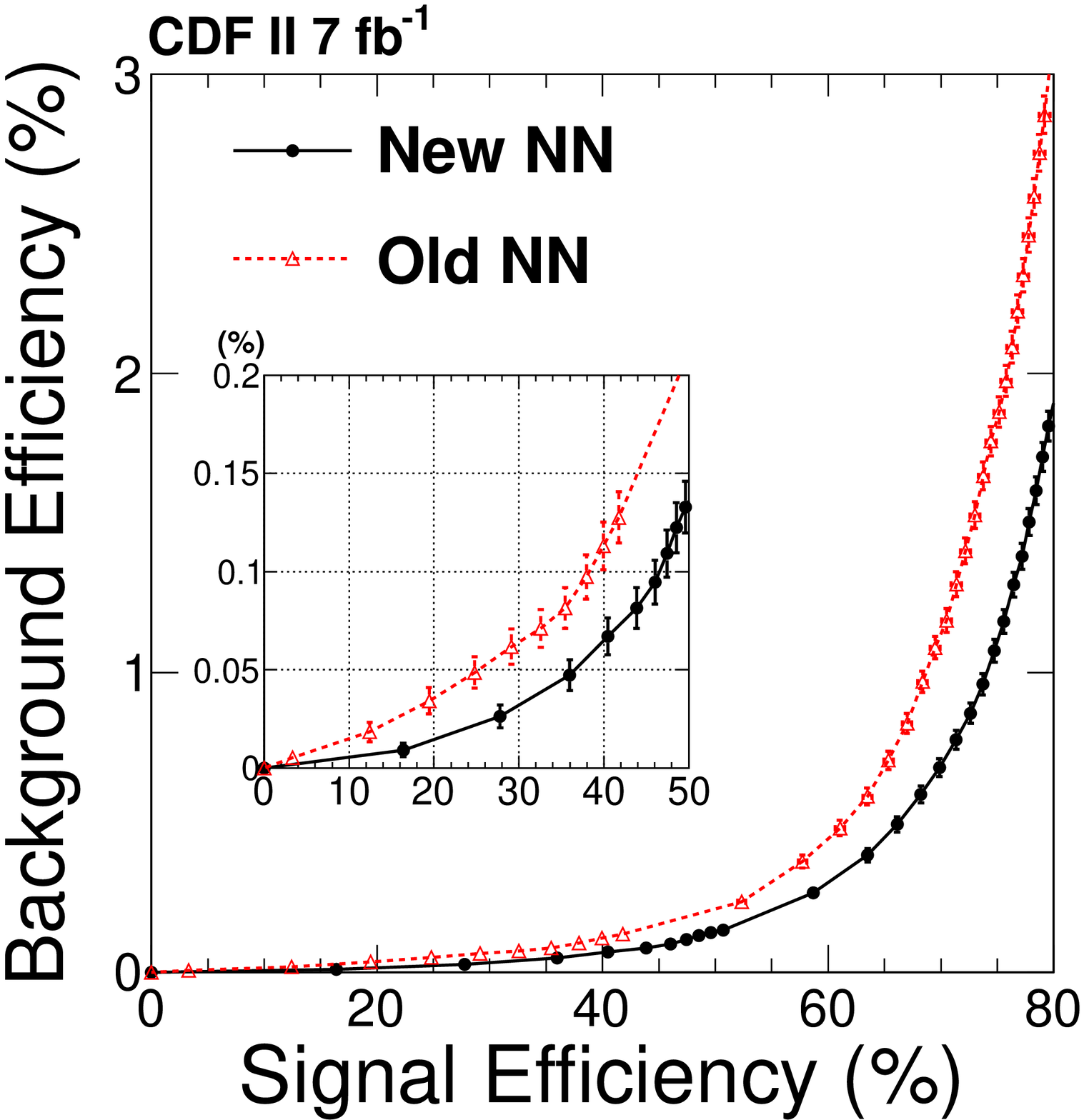}
  \caption{\textbf{Left:} CDF Muon detector layout. The blue region has been added for the updated analysis.
    \textbf{Right:} Comparison of signal and background efficiency for new and old NN.}
  \label{fig:miniSkirtsNNOut}
\end{figure}

CDF has updated their search with an improved analysis and significantly more (+3.2 \fb) data. 
The forward muon acceptance (left Figure~\ref{fig:miniSkirtsNNOut}) 
has been increased and the neural network has been improved to achieve twice the background rejection (right Figure~\ref{fig:miniSkirtsNNOut}). Additionally
the neural network has been extensively tested for mass bias. The peaking background prediction has
also been improved. The central-forward (CF) channel of the analysis has increased in statistics by $\sim$15\% resulting
in a total increase for both channels of $\sim$7\%.

\begin{figure}[htb]
  \centering
  \includegraphics[height=1.65in]{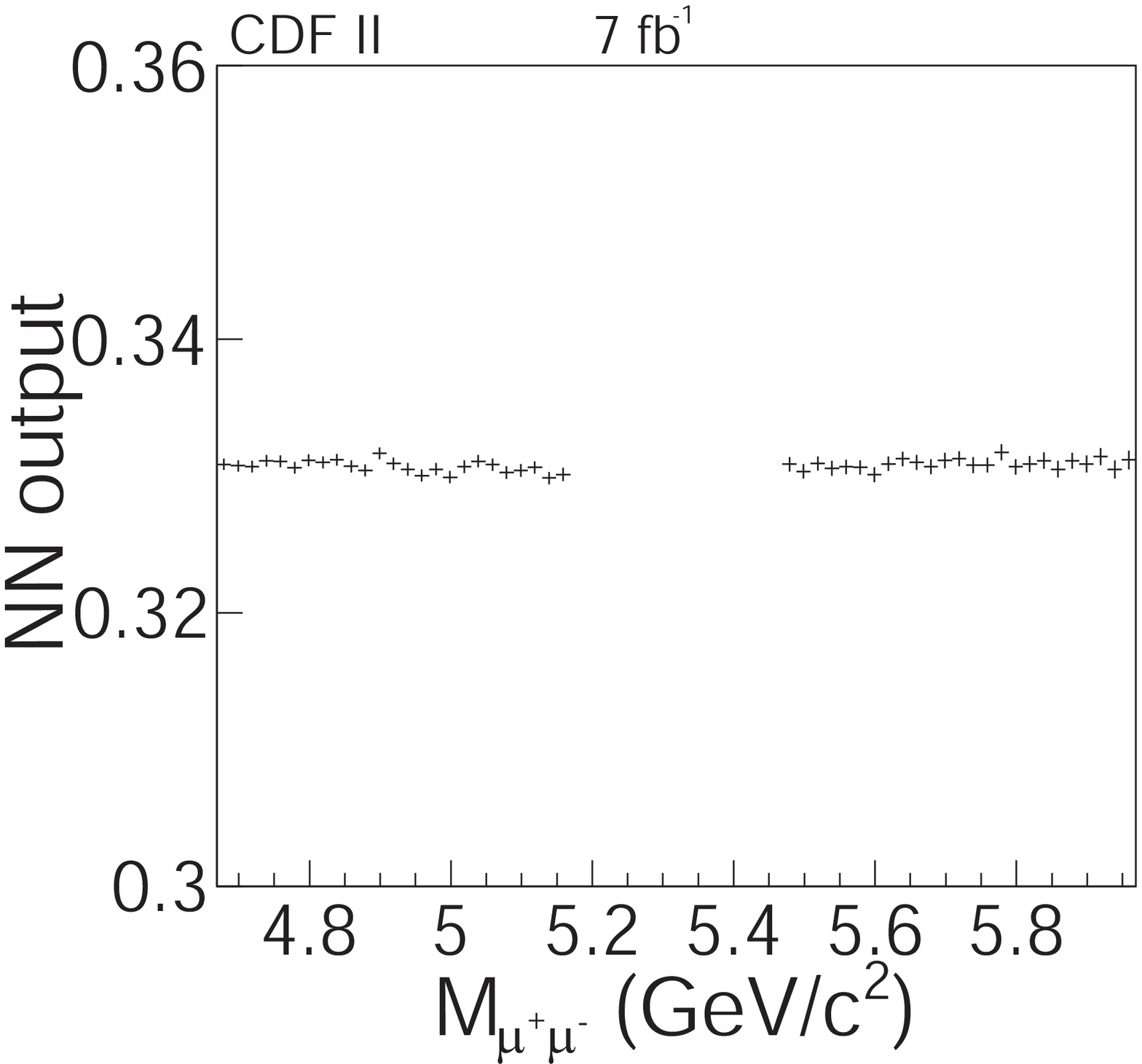}
  \includegraphics[height=1.65in]{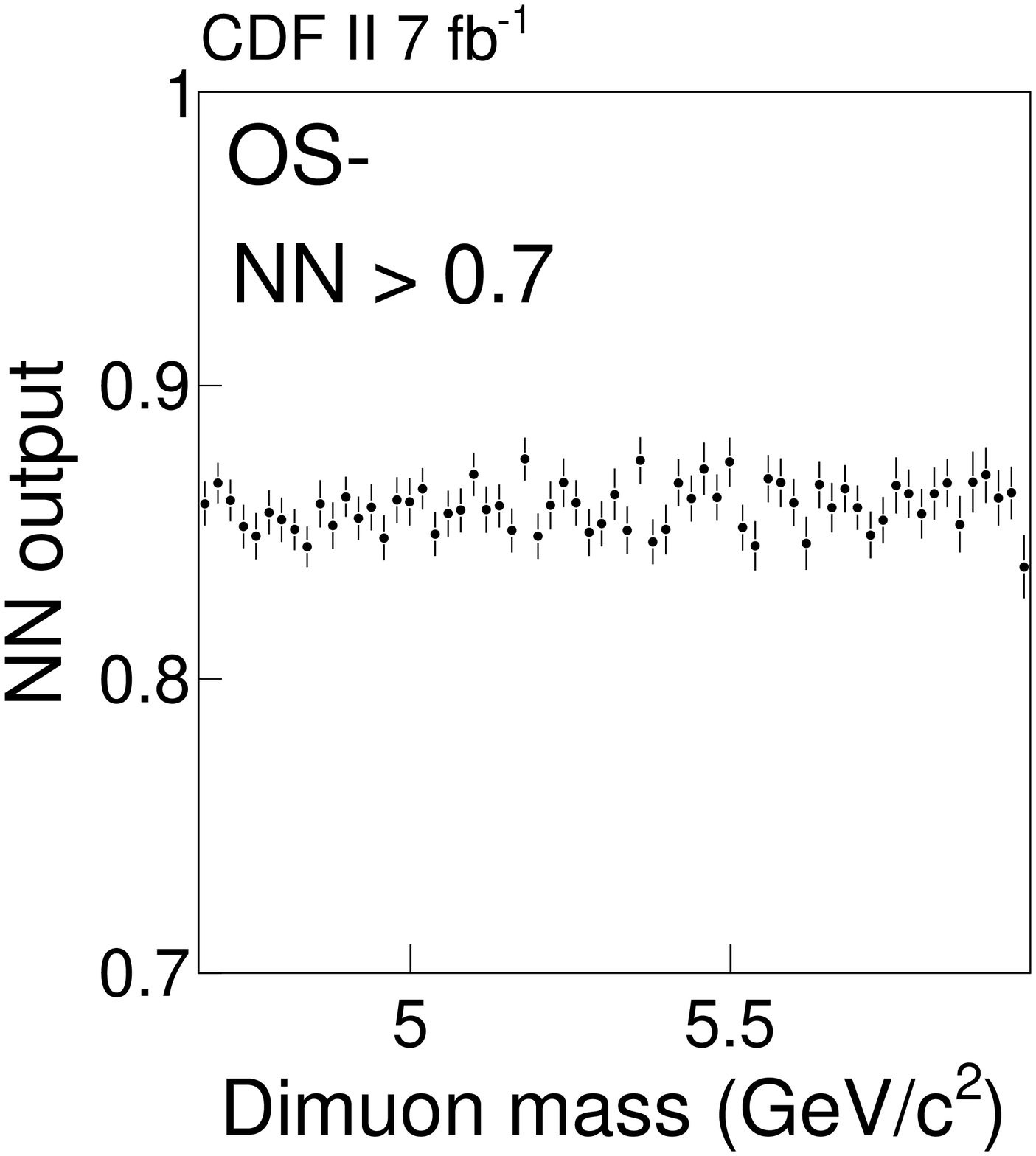}
  \caption{\textbf{Left:} NN output as a function of di-muon mass for the signal sample.
    \textbf{Right:} NN output as a function of di-muon mass for the OS- control sample (see Sect.~\ref{sec:control}).}
  \label{fig:backsculpt}
\end{figure}

The final NN network consists of 14 input variables.  Extensive cross checks were done to demonstrate the 14-variable network does not sculpt 
the di-muon invariant mass distribution. Figure~\ref{fig:backsculpt} shows the correlation across the \Mmm\ mass range.  No significant correlations are seen. 

\subsection{Control Samples for Background Estimates}
\label{sec:control}

The background estimate procedure is cross checked on four independent background samples:
\begin{description}
\item[OS-:]  opposite-sign muon pairs, which pass the baseline and vertex cuts
  and have a negative lifetime;
\item[SS+:]  same-sign muon pairs, which pass looser baseline and vertex cuts
  and have a positive lifetime;
\item[SS-:] same-sign muon pairs, which pass looser baseline and vertex cuts 
  and have a negative lifetime;
\item[FM+-:]  opposite-sign fake-muon pairs, at least one leg of which is 
  required to {\it{fail}} the muon ID requirement, passing 
  looser baseline and vertex cuts for both positive and negative lifetimes.
\end{description}

These are representative of various background contributions and thus are excellent control regions
to test the method of background estimation.
The estimated backgrounds are compared with actual observation. 
The agreement between predicted and observed number of background events in the signal region is good across all 
control regions.

\subsection{Results}

\begin{figure}
  \centering
  \includegraphics[width=0.39\textwidth]{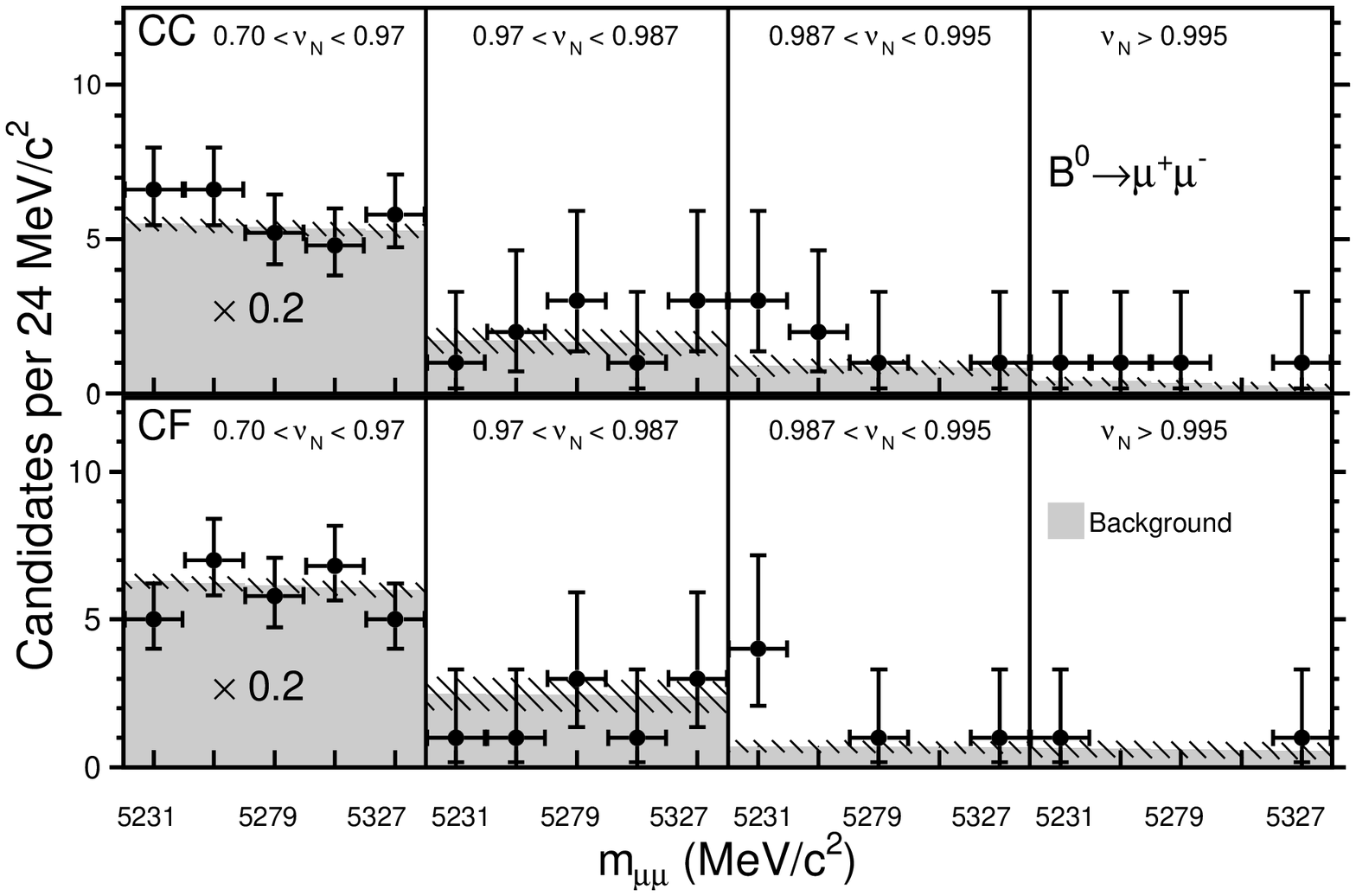}\\
  \includegraphics[width=0.39\textwidth]{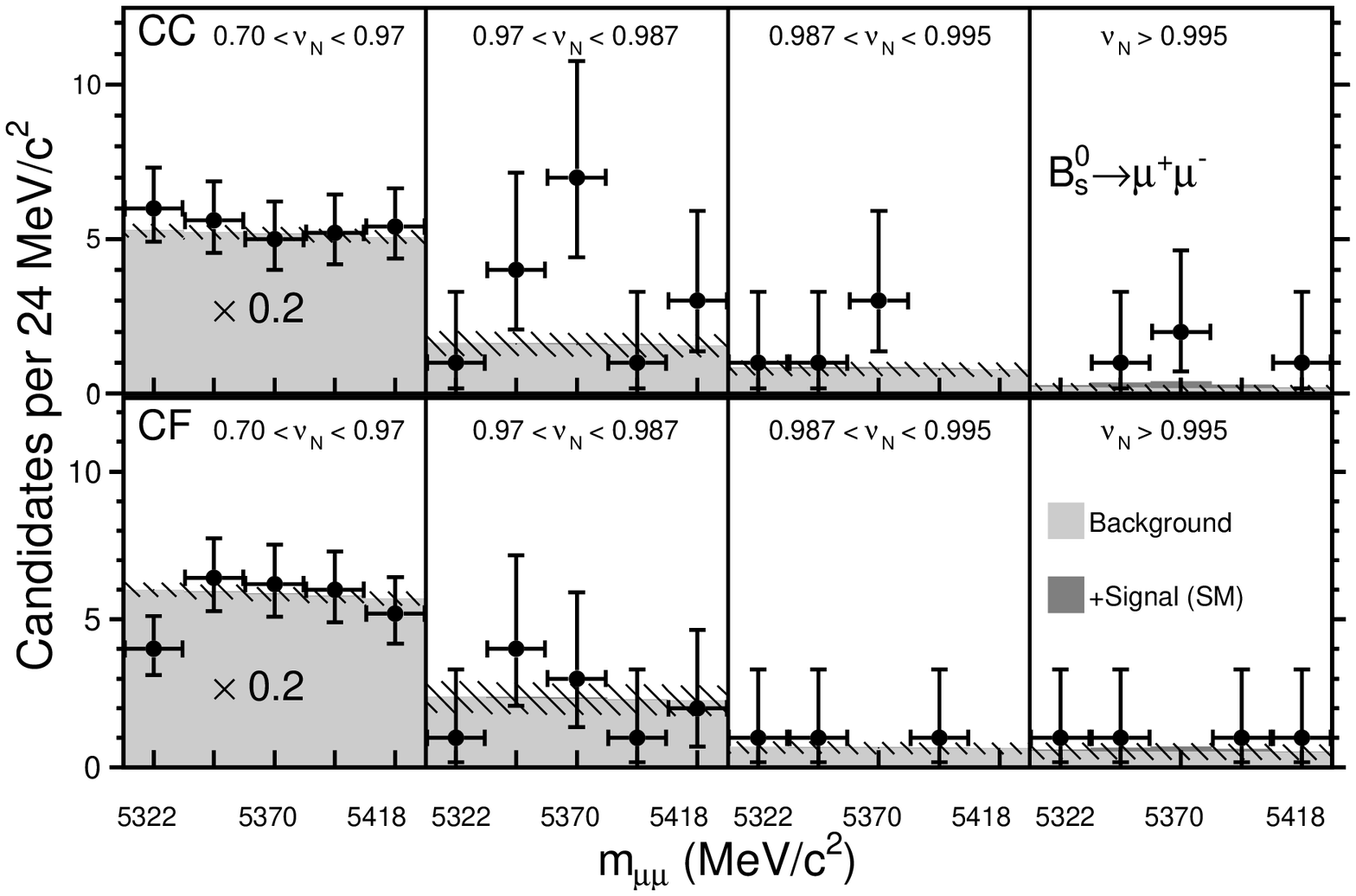}
  \caption{\textbf{Top:} Background estimates (light gray), systematic error on background estimates (hashed area), Poisson error on the mean
(error bars on points), and data for the \bd\ signal window.
\textbf{Bottom:} Similar plots for \bs\ with the addition of the SM expectations (dark gray). The data is divided into 8 NN bins, of which lowest 5 NN bins are combined into one bin for both figure, 5 mass bins, and two muon topologies (CC and CF).}
  \label{fig:resultsPlots}
\end{figure}

The observed events for the \bdmm\ search are shown in the top part of Figure~\ref{fig:resultsPlots}.  
The data is consistent with the background prediction and yields 
an observed limit of $\brbdmm < 6.0$ $(5.0) \times 10^{-9}$ at 95\% (90\%) C.L. 

An ensemble of background-only pseudo-experiments are employed to
estimate the significance as a $p$-value. The effects of systematic uncertainties are included in the
pseudo-experiments by allowing them to float within Gaussian constraints.
The resulting background-only $p$-value for the \bdmm search is 23.5\%. 

In the \bs\ search region the data exceed the background prediction in bins with $NN > 0.97$.
The bottom part of Figure~\ref{fig:resultsPlots} 
contains the detailed breakdown of the expected background and actual observation for the individual NN and mass bins for the \bsmm\ 
search. 

The source of the data excess in the $0.970<NN<0.987$ bin of the \bs\ signal region was investigated. The
same events, same fits, and same methodologies are used
for both the \bs\ and \bd\ searches. Because the data in
the \bd\ search region shows no excess, problems with
the background estimates are ruled out. 
Problems with the NN are ruled out by the many studies performed. The
most plausible remaining explanation is that this is a statistical fluctuation.

The $p$-value for the background-only hypothesis is 0.27\%.
Additionally a $p$-value of 1.92\% is also calculated assuming the SM+background hypothesis.

A $\Delta\chi^2$ fit is used to determine the \brbsmm\
most consistent with the data in the \bs\ search region.
A central value and 68\% C.L. is calculated at
\brbsmm=$(1.8^{+1.1}_{-0.9})\times 10^{-8}$. 
Additionally a bound at 90\% (95\%) C.L. on the branching fraction of \bsmm\ is set at $ 5.4 \times 10^{-9} < \brbsmm < 3.9 \times 10^{-8}$
($2.8 \times 10^{-9} < \brbsmm < 4.4 \times 10^{-8}$). 


Finally, upper limits at 95\% (90\%) C.L. of $\brbsmm < \brbsmmo $  $(3.5 \times 10^{-8})$ are set with the CLs methodology. 

\section{Conclusion}
\label{sec:Conclusion}
FCNC decays are a powerful probe for New Physics and CDF continues to lead
the searches in the B sector. The \bsdmm\ analysis has yielded the first two sided bound to \brbsmm\ and will be updated 
with the full CDF Run II data set. CDF also has the first observation of \lambdabmmlambda\
as well as angular measurements that compete and agrees with the B-factory results. \AT\ and \AIm\
are also measured for the first time for \bmmksComb\ decays. 


\end{document}